\documentstyle[fleqn]{article}
\begin{document}

\begin{center}
{\Large PP - WAVES}\\[4mm]
{\large The original article: Physical Review Letters {\bf3}, 571 
(1959)} \\[6mm] \rule{12cm}{.3mm}\\[6mm]
SOME GRAVITATIONAL WAVES$^*$\\[3mm]  Asher Peres\\
{\small Department of Physics, Israel Institute of Technology, Haifa,
Israel\\ (Received October 19, 1959)}\end{center}\medskip

The few hitherto known wave-like exact solutions of the Einstein
gravitational equations represented either plane$^{1,2}$ or
cylindrical$^3$ waves. We here intend to derive a new class of
solutions, displaying a lesser degree of symmetry, and thus more
generality.

Let us consider the metric
\[ ds^2=-\,dx^2-dy^2-dz^2+dt^2-2f(x,y,z+t)(dz+dt)^2.\]
One can easily show that $R_{zz}=R_{zt}=R_{tt}=f_{xx}+f_{yy},$
and other components vanish. If $f$ is a harmonic function of $x$ and
$y$, this metric thus satisfies Einstein's equations in vacuo, whatever
may be its dependence on $(z+t)$.

A convenient tetrad of orthonormal vectors is
\[ h_{(1)}{}^k=(\cos a,\sin a,0,0),\]
\[ h_{(2)}{}^k=(-\sin a,\cos a,0,0),\]
\[ h_{(3)}{}^k=(0,0,1-f,f),\]
\[ h_{(4)}{}^k=(0,0,-f,1+f),\]
where $\tan(2a)=(f_{xy}/f_{xx})$. The only nonvanishing independent
physical components$^4$ of the curvature tensor are
\[ \sigma=R_{(m1n1)}=-R_{(m2n2)}=(f_{xx}{}^2+f_{xy}{}^2)^{1/2},\]
where $m$ and $n$ take the values 3 and 4 only. Our metric thus belongs
to the second class of Petrov's classification.$^4$

Let us examine two examples. If
\[ f=(x^2-y^2)\sin(z+t),\]
one has $\sigma=2\sin(z+t)$. This is a plane monochromatic wave.

Another example is
\[ f=xy(x^2+y^2)^{-2}\exp[b^2-(z+t)^2]^{-2}\quad\mbox{for}\quad
  |z+t|<b,\]
\[ f=0\quad\mbox{for}\quad|z+t|\geq b.\]
(Note that $f$ has derivatives of all orders at $|z+t|=b$.) This is a
wave packet travelling with unit velocity in the negative $z$ direction.
It contains the singular segment $x=y=0$, $-b-t<z<b-t$. The question of
the existence of stable regular gravitational wave packets is still
open.

If $f_{xx}+f_{yy}\neq0$, one still has $g^{mn}R_{mn}=0$ and 
$g^{mn}R_{mr}R_{ns}=0$. The source of $f$ can then be interpreted as a
null electromagnetic field.$^5$

\bigskip\small
$^*$Partly supported by the U. S. Air Force, through the Air Research
and Development Command.

$^1$H. Bondi, Nature {\bf179}, 1072 (1957); H. Bondi, F. A. E. Pirani,
and I. Robinson, Proc. Roy. Soc. (London) A {\bf251}, 519 (1959).

$^2$H. Takeno, Tensor {\bf7}, 97 (1957); {\bf8}, 59 (1958), and {\bf9},
76 (1959).

$^3$N. Rosen, Bull. Research Council Israel {\bf3}, 328 (1954).

$^4$F. A. E. Pirani, Phys. Rev. {\bf105}, 1089 (1957).

$^5$C. W. Misner and J. A. Wheeler, Ann. Phys. {\bf2}, 525 (1957).
\end{document}